\documentclass[preprint, 10pt]{elsarticle}
\usepackage{rotating}
\usepackage{multirow} 
\usepackage{amstext}
\usepackage{amssymb}
\usepackage{graphicx}
\usepackage{amsmath}
\usepackage{multicol} 
\usepackage{listings}
\usepackage{pdfpages}
\usepackage{subcaption}
\usepackage{array}
\newcolumntype{M}{>{\centering\arraybackslash}m{\dimexpr.25\linewidth-2\tabcolsep}}

\begin{document}

\begin{frontmatter}

\title{Predicting encounter and colocation events in metropolitan areas}


\author[unimi]{Karim Karamat Jahromi, Matteo Zignani, Sabrina Gaito and Gian Paolo Rossi }

\address[unimi]{Computer Science Department, University of Milan, Italy}

\begin{abstract}

Despite an extensive literature has been devoted to mine and model mobility features, forecasting where, when and whom people will encounter/colocate still deserve further research efforts. Forecasting people's encounter and colocation features is the key point for the success of many applications ranging from epidemiology to the design of new networking paradigms and services such as delay tolerant and opportunistic networks.
While many algorithms which rely on both mobility and social information have been proposed, we propose a novel encounter and colocation predictive model which predicts user's encounter and colocation events and their features by exploiting the  
spatio-temporal regularity in the history of these events. 
We adopt a weighted features Bayesian predictor and evaluate its accuracy on two large scales WiFi and cellular datasets. Results show that our approach could improve prediction accuracy w.r.t standard na\"{i}ve Bayesian and some of the state-of-the-art predictors.

\end{abstract}


\end{frontmatter}
\captionsetup{subrefformat=parens}

\section{Introduction}\label{sec_intro}

Mobile phones are desired options for tracking and mining user behavior in daily life since they are usually carried and placed in close proximity to the users.
Smartphones can unobtrusively be used to collect data about people's behavior in multiple aspects. These include people's mobility, i.e. their places, how they move among places and whom they could meet while spending time in those places. The study of human mobility has long been a hot topic for research in the last decade. We know, for instance, that the mobility of individuals is not a random process, but, by contrast, it shows a high level of spatial and temporal regularities when observed on a daily time frame as a consequence of the pace dictated by professional responsibilities and social behaviors \cite{Barbosa2015,Jia2012,Hasan2013,Hsu2010}.
Although an extensive literature has been devoted to mine mobility features and create accurate mobility models, it still represents an open research problem and especially the ability to predict the encounter or colocation events is a task still deserving further research efforts.
Forecasting the occurrence of this kind of events among mobile careers can be utilized in delay tolerant and opportunistic networks and may lead to achieve high efficiency in performing routing and data forwarding activities \cite{Karamshuk2011,Ciobanu2012,Chilipirea2013}. In a scenario of high dynamics and intermittent radio connectivity the awareness about the approximate location, the time duration of an encounter or colocation event and people involved goes further system implications, paving the way for novel applications in a variety of fields, including commercial ADs, recommendation systems, and mobile social networks. 
A few location prediction algorithms have recently achieved accurate results \cite{Wang2011,Zhang2015} by combining information about individuals' mobility patterns and social ties and behavior. However, data about human sociality are difficult to be collected and it will become even harder to obtain them in the future as a consequence of the tightening of restrictions regarding privacy preservation. In contrast, mobility patterns will remain an accessible information because users may be willing to provide it by autonomously enabling geo-localization on their mobile device. In this scenario, one of the challenges today is the design of algorithms to predict where, when and with whom a user will experience an encounter or colocation event by simply leveraging the mobility patterns of the user, especially where knowing people's encounter and colocation events is the key point for the success of the applications and protocols. When relying on spatial-temporal information only, a few algorithms were designed to predict the next place visited by a user with good accuracy \cite{Scellato2011,Trasarti2015,Do2015}; however, to the best of our knowledge, only in \cite{Ciobanu2012} the issue of predicting the next encounter has been addressed.

In this work, we use spatio-temporal mobility information to design a novel algorithm able to predict with high accuracy the next encounter or colocation event along with its characteristics, such as location, duration, and people involved. Specifically, the algorithm learns patterns from people's mobility and their encounter/colocation history and predicts the next encounter and colocation events by exploiting weighted features Bayesian predictor. The approach has been extensively evaluated on two large datasets, covering different real scenarios and mobility settings.

\section{RelatedWork}\label{sec_relatedwork}

Pattern recognition and prediction are closely related tasks since human movement cues are usually periodic and/or repetitive \cite{Clauset2007,Helmy2011,Kim2007}. Therefore repetitive encounter and colocation events can be learned and predicted reliably as long as we could collect enough observation data from smartphones and other mobile devices.

One challenge for using the WiFi and Cellular datasets for prediction (especially PoI prediction) is that PoIs are represented by ID (symbolic place) without any coordinates. Therefore some existing prediction scheme such as \cite{Monreal2009,Scellato2011,Trasarti2015}  would not be applicable. So our used datasets do not support the arithmetic or logic operation, which is usually are used to process GPS coordinates for location prediction.

Peddemors et al. \cite{Peddemors2010} propose an approach based on the prediction of the time of next occurrence of an event of interest, such as arrival time to a certain place with a focus on the prediction of network visibility events as observed through the wireless network interfaces of mobile devices. Their approach is based on a predictor that analyses the events stream for forecasting context changes. The authors found that including predictors of infrequently occurring events can improve the prediction performance.    

Gao et al. in \cite{Gao2012} proposed a location predictor model that captures the spatio-temporal contexts of the visited places. They exploited a smoothing technique in the training of spatio-temporal model, to avoid the over-fitting problem due to a large number of spatio-temporal trajectory patterns. They assume that temporal features (day of the week and hour of the visit of a place) to be independent, 
and estimate the distributions of the day of the week and the visiting hour by Gaussian distributions. 

In contrast to the wide range of future place prediction works \cite{Song2006,Asahara2011,Gambs2010,Lu2013,Gomes2013} relying on Markov chain, needed to keep track n previous visited places, 
our proposed approach just need temporal context as an inquiry for predicting PoIs.

The Next Place predictor method in \cite{Scellato2011} captures the concurrent temporal periodicity of mobile users when they visit their most important places. This spatio-temporal predictor relies on a non-linear time series analysis of the arrival time and on pause time durations of users in their most relevant places. This predictor, besides predicting the arrival time to the next place and its stay duration, is also able to  predict the interval time between two subsequent visits to the predicted place. This approach has been only applied to the most important visited places and needs a large amount of data to constitute time series. Due to these requirements, its application is limited just to the most frequently visited places (home and workplaces). 

In \cite{Do2015} a probabilistic kernel method for visited place prediction using spatio-temporal information via multiple kernel functions are presented. The kernel density estimation is a smoothing technique for sparse data collected by smartphones. 
Even though this approach exploits i.i.d assumption among spatio-temporal context in Bayesian predictor, it has obtained good accuracy in predicting the next visiting places just for the next few hours.

Authors in \cite{Zhang2015}, by analyzing MIT Reality Mining CDR dataset, have observed a strong correlation between calling pattern and colocation patterns of mobile users. By exploiting this social interplay on top of user periodic behaviors, they proposed a self-adjuster symbolic predictor which combines the output of social interplay and periodicity predictors to estimate the next cell to visit. 
Although authors only used calls pattern, they achieve higher prediction accuracy than the other state of the art schemes at cell tower level. Considering that MIT Reality Mining CDR has been collected in 2004 and during that period definitely calls were dominated contact activities among people, so it makes sense if authors exploited call activities for capturing social interplay among participants in their experiment while nowadays the majority of contact actives among people and friends have been oriented towards the wide variety of Internet-based applications. However, such Internet-based contact activities somehow will be hidden from CDR datasets, since when a user accesses to the Internet just the Internet traffic data will be recorded in CDR. Therefore nowadays we should be  conservative about extracting social interplay among mobile users just relying on calls and even text SMS. 

Authors in \cite{Ciobanu2012,Ciobanu2015} according to their observations of the time of encounters (nodes encounter patterns tend to have centers in the middle of the day), proposed the Gaussian process for predicting the time of encounters and also predicting the number of encounters in the predicted time frame.

In other works \cite{Nguyen2012,Ciobanu2015} authors predicted the user future contacts to be used in a routing protocol to improve the efficiency of message delivery in opportunistic networks.

\section{Datasets}\label{sec_Dataset}

To validate the proposed approach, in this paper we adopt two different datasets each covering a different mobility scenario. The first dataset is WiFi and has been collected through Access Points (APs) in the Dartmouth university campus \cite{Kotz2013}. Whenever a mobile device (smartphone, tablet or laptop computer) associates or disassociates to an AP, a log message is recorded. Each record contains a timestamp in seconds, the MAC addresses of the AP and of the mobile device, the Access Session Time in seconds, and the Access Session Status (Start - attach, or Stop - detach). The Dartmouth WiFi dataset \cite{Kosta2014} lasts 4 months, from January 3rd to April 30th, 2004, and contains mobility patterns of 17414 anonymized mobile users and 1292 APs, whose coordinates are not provided.
The mobility trace of a user $u$ is represented as a sequence of Points of Interest (PoIs) \cite{KJ2015} visited by $u$ and temporally annotated with his arrival and departure times.
In order to extract significant PoIs from the WiFi dataset, we filtered out APs where the user just passed through by considering only APs visited by the user for more than 15 min. Under this condition, the WiFi dataset still includes 14082 mobile users and 907 APs (PoIs).


The second dataset is a large anonymized dataset of Call Detail Records (CDRs) containing voice, text, and data phone activities of nearly 1 million mobile subscribers. The records, provided by one of the Italian mobile operators, rely on activities gathered for a total of 67 days, from March 26th to May 31st, 2012, in the Milan metropolitan area. 
Whenever a voice call or text or data activity is issued, a CDR is created to record calling and called user-ID, date, time, location, and, in the case of a voice call, its duration in seconds. 
The location is expressed in terms of cell tower ID and its location-name attribute, e.g. street/square name or city's zone, that represents a coarse grain division of the city region. 
The entire dataset contains more than 69 million phone-call records and 20 million text message records.

\begin{figure}
	\centering
	\includegraphics[width=1\textwidth]{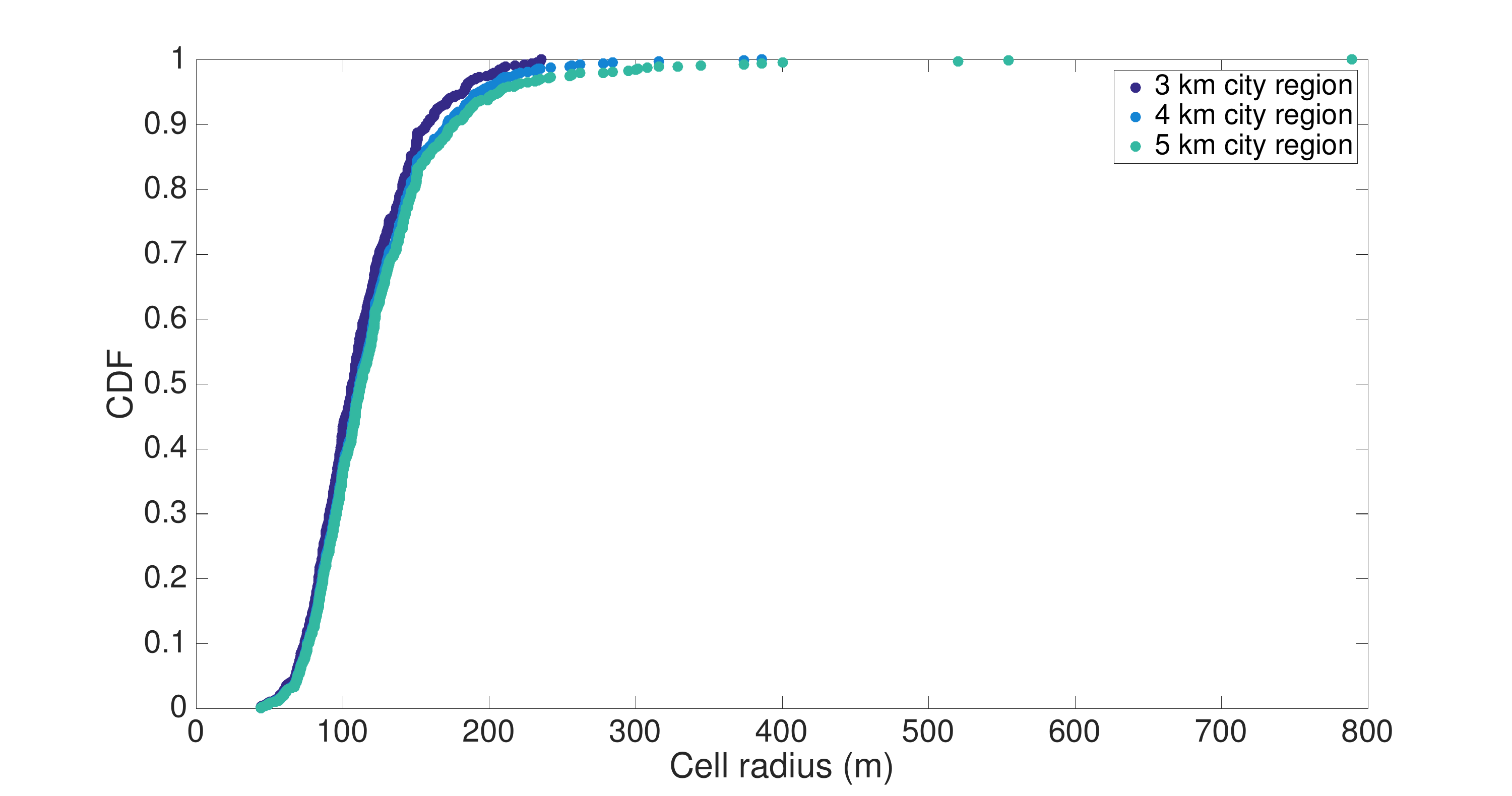}
	\caption{Distribution of the radius of the cell towers in the Milan area.}
	\label{fig:tower_radius}
\end{figure}

Cellular network datasets are very rich sources of information for studying and analyzing human mobility. Nonetheless, they often raise concerns about their low temporal and spatial precision \cite{Csáji13,Gonzalez2008, Isaac11}. Temporal concerns are easily sidestepped by considering data traffic that allows a finer temporal analysis then voice/text phone activities. Spatially, CDR accuracy is constrained by the coverage area of cell towers, which varies from a few hundred square meters in urban areas to a few square kilometers in rural areas. We performed the spatial analysis of our urban dataset, see \cite{QuadriSmartcomp2016}, and we obtained the cumulative distribution function of the radius of the cell towers for the areas of the city at the different distance from the center.  Figure \ref{fig:tower_radius} depicts that the median radius value is of some 120m in the inner circle of the city (within 3 km from the center).

According to this result, PoIs correspond to cell towers and its coverage area. Although the coarse spatial granularity of the cellular towers better fits the definition of the region of interest, in this paper we mainly use the term Point of Interest to make the notation and the presentation more uniform.

One challenge when having at disposal WiFi data or CDR without spatial coordinates is represented by the impossibility to apply mobility prediction schemes such as \cite{ Monreal2009,Scellato2011,Trasarti2015}  which are based on GPS information.

\section{Encounter and Colocation Events}\label{sec_EncounterEvent}

An encounter event means meeting face to face, which implies physical proximity among people. The extent of this physical proximity is not always exactly clear and may be vary on different scenarios, applications, and domains. For instance, in the biological field and in disease spreading, physical proximity is short, while in wireless networks it depends on the coverage areas of mobile devices or wireless network infrastructures. Nowadays smartphones are so widely carried by humans that can be used to observe mobility and extract physical proximity information. In the communication network literature, an encounter among mobile devices occurs when they are in the communication range or when they are within the same coverage area of the communication network infrastructure, the latter also called "indirect encounter" \cite{Zhao2008,Legendre2011}. Although this definition may not always reflect proper and exact realistic physical encounters among mobile nodes due to some challenges\cite{Jahromi2014,Jahromi2015}, 
most researchers define an encounter event occurrence in a WLAN when two or more mobile nodes are associated to the same AP during an overlapped time interval. 
Despite some challenges and limitations, if collected WiFi datasets are used carefully (i.e. accounting for the effects of ping-pong events, overlap in coverage areas and missed encounters) it would appear to be a good source of empirically-derived data on human encounters since large amount of data can be gathered easily at low cost, allowing even large-scale analysis of encounter patterns.
Here for WiFi dataset, we use smoothing the ping-pong events according to \cite{Jahromi2014}, for extracting encounter events.

The resulting record for an encounter event is:
\begin{verbatim}
UserA,UserB,PoI Id,Encounter Start Time,Encounter End Time.
\end{verbatim}

In an almost similar way, colocation event has been defined in cellular networks among mobile users while they are connected to the same cell tower for an overlap time interval. Taking into account the coarser temporal and spatial granularity of CDR datasets w.r.t. WiFi dataset, therefore there is a significant difference in the spatial range of encounter and colocation events. Each colocation event is characterized by a specific time interval and place (PoI). Authors in \cite{Zhang2015,Calabrese2011} have characterized the spatio-temporal features of colocation events and observed a reasonable subset of actual face-to-face meetings between users.
To estimate the colocation events, mobile users are assumed to stand under the coverage area of the same cell for a time period lasting $T_{h}$ seconds before and after each on-phone activity. For instance, if $t_{u}^{k}$ and $Td_{u}^{k}$ are the initial time stamp and the call duration of the $k$-the activity of user $u$ in cell $c$, then we assume that the mobile user $u$ is available under the cell $c$, at least within the time interval [$t_{u}^{k}$-$T_{h}$ , $T_{h}$+$t_{u}^{k}$+$Td_{u}^{k}$]. 
For messaging activities $Td_{u}^{k}=0$ holds and in general we set $T_{h}=900$ seconds for extracting colocation events from CDR dataset.

The resulting record for a colocation event is:
\begin{verbatim}
UserA,UserB,PoI Id, Colocation Start Time, Colocation End Time.
\end{verbatim}

\section{Encounter/Colocation Prediction Methodology}\label{sec_EncounterPrediction}

While predicting encounter or colocation events, we seek the answers to three questions \cite{Pirozmand2014} : 1) where will the encounter and colocation occur, given a user and a future time? 2) how long will s/he be with other users at that PoI, i.e. the event duration? and 3) whom will s/he meet (i.e., encounter/colocation contacts)?
In this work, our goal is to predict the places (PoIs) where a user will experience an encounter or colocation event, give an estimate of its duration and which people are involved.

We assume no a priori knowledge on the temporal relation between encounter or colocation events. Focusing on the temporal and the spatial information of the encounter/colocation traces, we learn the dependencies between these contextual variables and next events. 
The temporal context captures regular patterns in the occurrence of events from the weekly calendar, such as events occurring at given time of day and day of the week. On the other hand, the dynamics of these events can be explored through the spatial information.
Since daily schedule of people usually is different on weekdays and weekend, to construct the encounter or colocation predictors, we consider several parameters to efficiently capture multiple aspects of the temporal contexts.
The temporal context features are: \textit{i}) day time slot, and \textit{ii}) day of the week. The "day time slot" $\iota$ is an integer feature and depends on the length $H$ of the time slot, i.e $\iota=\left \lfloor t/H \right \rfloor $ where $t=0,\ldots,23$. We set $H=2$ hours since it represents a trade-off which offers a good daily resolution and a robustness against small changes in the daily movement routine, e.g. being late for work due to an exceptional traffic jam or little delays in the agenda.

The "day of the week" $\phi$ maps a day of the week to an integer, where Monday is 1 and Sunday is 7.
We compute the above features on the encounter and colocation start and end times so that each encounter or colocation record is defined by the user-ids of the mobile users, the PoI IDs, $\phi$, $\iota_{s}$ for the start time and $\iota_{e}$ for the end time (see Table \ref{tab:input_format}).

\begin{table}[]
	\centering
	\caption{Encounter/Colocation features.}
	\label{tab:input_format}
	\begin{tabular}{|l|l|l|l|l|l|}
		\hline
		\multirow{2}{*}{$\phi$} & \multicolumn{2}{l|}{$\iota$} & \multirow{2}{*}{UserID} & \multirow{2}{*}{PoIID} & \multirow{2}{*}{Encountered/Colocated User ID} \\ \cline{2-3}
		& $\iota_{s}$ & $\iota_{e}$ &  &  &  \\ \hline
	\end{tabular}
\end{table}

We adopt a per-user perspective, i.e. for each mobile user $u$ we will predict the encounter and colocation characteristics relying on her/is context history, only. The predictor, trained on $u$'s event records, will accept as input the tuple 
$T=(\phi,\iota)$  and will return the PoI where the encounter or colocation will occur, its duration and the users encountered or colocated by $u$ during the temporal context $T$.

\section{Predictive Model}\label{sec_ProbabilisticModel}
The na\"{i}ve Bayesian classifier is one of the most common classification techniques.
Na\"{i}ve Bayesian classifiers are based on the Bayes' theorem with na\"{i}ve independence assumption between the features and apply a decision rule, known as Max a Posteriori or MAP decision rule, which selects the hypothesis/class with the highest probability. In this work, similarly to other recent mobility prediction works \cite{Do2015,Gomes2013,Etter2013,Chilipirea2013}, we use Bayesian classifier for encounter and colocation prediction.
Beside its simplicity and being fast compared to other classifiers, Bayesian classifier can be trained with a few observation records, especially in our case where encounter and colocation traces are sparse and sporadic, and still achieve reliable results.

\subsection{Encounter/Colocation PoI prediction}

Most of the recent location-based services are based on the knowledge of the current and future place of the mobile user. For instance, by exploiting the future visiting places, we can access to the information such as nearby PoIs or available services.
For the prediction of the encounter or colocation PoI, we consider the conditional probability of a PoI $L=l$ given the temporal context $T=(\phi,\iota)$.

Under independence assumption:

\begin{equation} \label{eq:bayes_simplify}
P( L=l\mid T) \propto P(L=l) P(\Phi=\phi \mid l) P(I=\iota \mid l)
\end{equation}

By exploiting the MAP decision rule we have:

\begin{equation} \label{eq.4}
\begin{array}{l}
l_{p}=
argmax_{j} ( P(\phi \mid l_{j}) P( \iota\mid l_{j}) P(L=l_{j}))
\end{array}
\end{equation}

The application of the standard formulation of a na\"{i}ve Bayesian classifier poses some problems due to the conditional independence assumption. In human mobility context people have a different schedule on weekdays and weekend, i.e a person may visit different places on weekdays and weekends during the same time slot. In this case, the independence assumption for $\phi$ and $\iota$ would be violated. To smooth the independence assumption, we used a na\"{i}ve Bayesian classifier with feature weights based on Kullback-Leibler divergence \cite{Lee2011}.\\
The Kullback-Liebler measures for feature $a$ and class label $c$ is defined as \\

\begin{equation} \label{eq.5}
\begin{array}{l}
\mathfrak{KL}(C\mid a)=\sum_{c}P\left (  c\mid a\right )log(\frac{P\left (  c\mid a\right )}{P(c)}) 
\end{array}
\end{equation}

Where $ \mathfrak{KL}(C\mid a)$ is the average mutual information between the class event $c$ and the feature value $a$ with expectation taken with respect to a posteriori probability distribution of $C$. This can be considered as an asymmetric information theoretic similarity between two probability distributions, which measures how dissimilar a priori and a posteriori. This distance measure corresponds to the amount of divergence between a priori distribution and a posteriori distribution. The weight of features can be defined as the weighted average of the $ \mathfrak{KL} $ across the feature values.

The introduction of the weights results in the following formulation of the predictor:
\begin{equation} \label{eq.6}
\begin{array}{l}
l_{p}=
argmax_{j} ( P(\phi\mid l_{j}) ^{w_{\phi,L}} P( \iota \mid l_{j}) ^{w_{\iota,L}} P(L=l_{j}))
\end{array}
\end{equation}

Where $w_{\phi,L}$ and $w_{\iota,L} $ 
are the feature weights of $\phi$ 
and $\iota$ calculated for PoIs label in training set according to \cite{Lee2011}. The weights are shared over all users in the training set.
Finally, since the na\"{i}ve Bayesian model returns the probability $P(L=l|T)$, we can retrieve the $k$-most likely places (PoI IDs) $l_{p}$ given the temporal context $T=(\phi, \iota)$.

\subsection{Encounter duration prediction}

The duration predictor estimates how long the encounter event at the predicted PoI will last. Indeed, the predictor depends not only on the temporal context $T$, but also on the outcome of the PoI predictor, i.e. $l_p$. In this setting we aim at finding the duration $d_j$ which maximizes $P(D|T,L=l_p)$, i.e.
\begin{equation} \label{eq.7}
d_{p}=argmax_{j}\left ( P(\phi,\iota,l=l_{p}\mid d_{j})P(D=d_{j}) \right )
\end{equation}

In above equation addition to temporal features, the predicted PoIs also considered as the spatial feature.
By applying the feature weighting for na\"{i}ve Bayesian classifiers we obtain:

\begin{equation} \label{eq.8}
\begin{array}{l}
d_{p} = argmax_{j} ( P(\phi\mid d_{j}) ^{w_{\phi,D}} P( \iota\mid d_{j}) ^{w_{\iota,D}}  P(L=l_{p}\mid d_{j}) ^{w_{l_p,D} } P(D=d_{j}))
\end{array}
\end{equation}

where $w_{\phi,D}$, and $w_{\iota,D}$ are the feature weights of $\phi$ and $\iota$, and $w_{l,D}$ is the weight related to the predicted PoI. So the event duration predictor will learn a function whose input is the tuple $(\phi,\iota, l=l_{p})$.

Encounter and colocation duration among mobile users in a specific place (PoI) varies in time. As a consequence, the prediction of the duration is not straightforward. On the other hand since most of the people follow daily schedule tasks, we expect that the variation of the duration lies in a limited range. These observations reflect on how we evaluate the accuracy for the encounter and colocation duration task. We extract from the test set $P_{z}$, the set of encounter or colocation durations occurring in the PoI $l_{j}$ for the temporal context $(\phi, \iota)$.   
After removing outlier durations by using skewness \cite{Heymann2012}; we obtain $P_{z}$=$\left \{ pt_{1},pt_{2},\cdots,pt_{\left |P_{z} \right |} \right \}$, where $\left|P_{z}\right|$ is the size of $P_{z}$. Then we compute the average $\mu_{z}$ and the standard deviation $\sigma_{z}$ on the set $ P_{z}$. If the predicted duration for the temporal context 
$\left \{\phi, \iota, l=l_{j} \right\}$ 
lies in the interval $\mu_{z}-\sigma_z,\mu_{z}+\sigma_{z}$, we consider the event duration prediction to be correct.

\subsection{Encounter/Colocation Contacts Prediction}

Because of the critical role of predicting future encounter or colocation events in content delivery and routing protocols in opportunistic and delay tolerant network \cite{Karamshuk2011,Vu2011},
in this section, we focus on predicting whom a user will meet in a specific period 
$(\phi, \iota)$.
In the Bayesian setting, it corresponds to find the user or the set of users maximizing the following probability:

\begin{equation} \label{eq.9}
c_{p}=argmax_{j}\left ( P(\phi,\iota\mid c_{j})P(C=c_{j}) \right )
\end{equation}

Since the set of people met by a mobile user may change between weekdays and weekend even during the same time slot, we alleviate the conditional independence assumptions by feature weighting:

\begin{equation} \label{eq.10}
\begin{array}{l}
c_{p}=
argmax_{j} ( P(\phi\mid c_{j}) ^{w_{\phi,C}}  P(\iota\mid c_{j}) ^{w_{\iota,C}} P(C=c_{j}))
\end{array}
\end{equation}

Where $w_{\phi,C}$ and $w_{\iota,C} $ are the feature weights of $\phi$ 
and $\iota$  are calculated for the contact label.

\section{Classifier evaluation}\label{sec:Evaluating Prediction Accuracy}

In this section, we evaluate the goodness of the Bayesian predictors with weighted features on the encounter and colocation traces separately due to the significant difference in the spatio-temporal granularity of WiFi and CDR datasets which results in the prominent difference in the spatial granularity range of encounter and colocation events.

In each subsection we train and evaluate the Bayesian classifier with weighted features and also the standard na\"{i}ve Bayesian classifier (NBC) for each user, separately. This way we obtain a set of accuracy values, whose distribution captures the performance of the approach for a specific task. Moreover, since the Bayesian classifier can return the most $k$ likely items, we report the results of the evaluation for $k=1,2,3$.
We conducted the evaluation by using 4-fold cross validation and the average accuracy as performance metric.

\subsection{Encounter prediction}

In encounter trace, users with at least 75 records are chosen for encounter prediction to have enough records for training the classifier. In this subsection, we evaluate the encounter prediction accuracy performance for PoIs, durations, and also contacts for input temporal query.

\subsubsection{Encounter PoI prediction}

\begin{figure}
	\centering
	\includegraphics[width=1\textwidth]{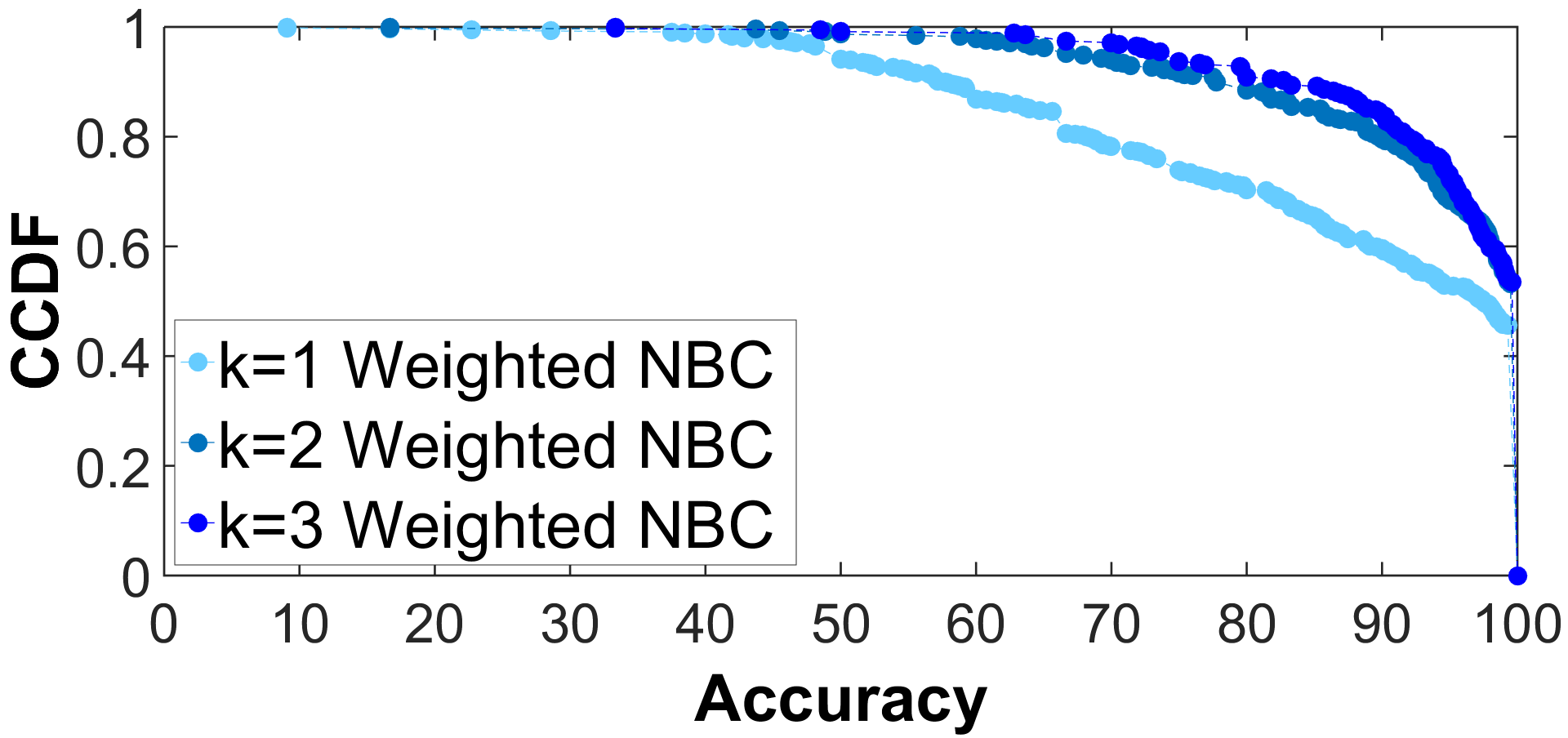}
	\caption{ The distributions of the encounter PoIs prediction accuracy for the Bayesian classifier with weighted features.}
	\label{fig:EncounterLocationDistribution-p-123-Weighted}
\end{figure}

In Figure \ref{fig:EncounterLocationDistribution-p-123-Weighted} we report the distribution of the accuracy for the encounter PoI prediction task and for $k=1,2,3$ through exploiting the weighted features Bayesian classifier. In this case (Figure \ref{fig:EncounterLocationDistribution-p-123-Weighted}), for $k=3$, more than 90\% of mobile users have more than 80\% accuracy in predicting PoIs where encounters will happen. Accuracy degrades for $k=1$, where around 70\% of predictors get more than 80\% accuracy. 

Finally, we observe that the accuracy for $k=3$ is higher than $k=2$ and $k=1$ cases since by increasing $k$ we enlarge the prediction set and the probability that the prediction set will contain the correct PoI.

\begin{figure}
	\centering
	\includegraphics[width=1\textwidth]{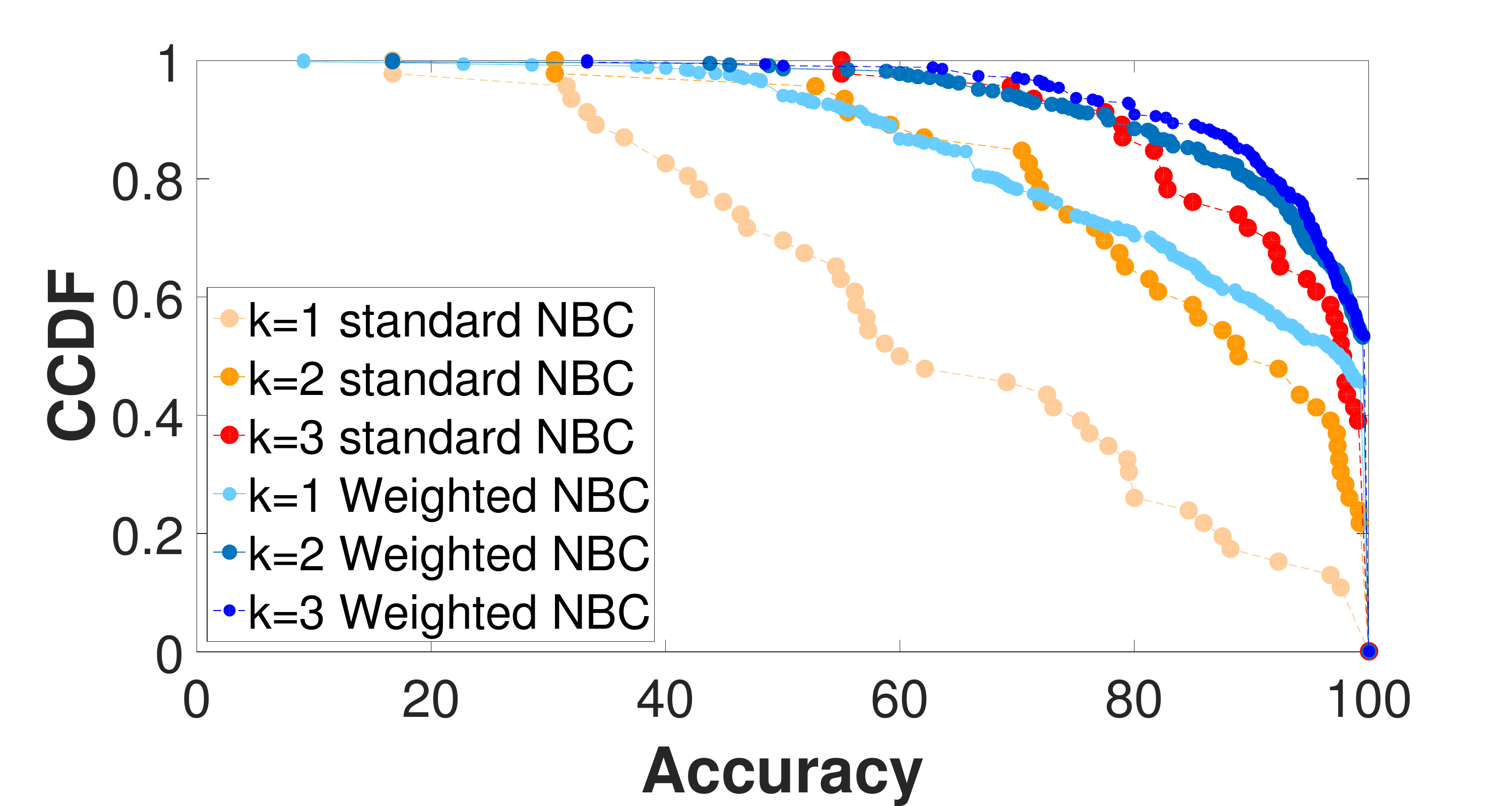}
	\caption{Comparison between the standard na\"{i}ve Bayesian classifier and its weighted version in terms of PoIs prediction accuracy distribution.}
	\label{fig:CompareLocationPredictionAccuracy}
\end{figure}

In section \ref{sec_ProbabilisticModel} we introduced the feature weighting to relax the conditional independence assumption. In the case of dependencies among the features we expect that the accuracy of the weighted features Bayesian classifier increases w.r.t. the standard formulation. To this aim in Figure \ref{fig:CompareLocationPredictionAccuracy} we compare the results between the standard na\"{i}ve Bayesian (NBC) and the  Bayesian classifier with weighted features in terms of PoI prediction accuracy distribution. 
We observe a pronounce improvement in the accuracy prediction, especially for $k=1$ and $k=2$, when we introduce feature weighting.

\begin{figure*}
	\centering
	\begin{subfigure}{0.48\textwidth}
		\includegraphics[width=\textwidth]{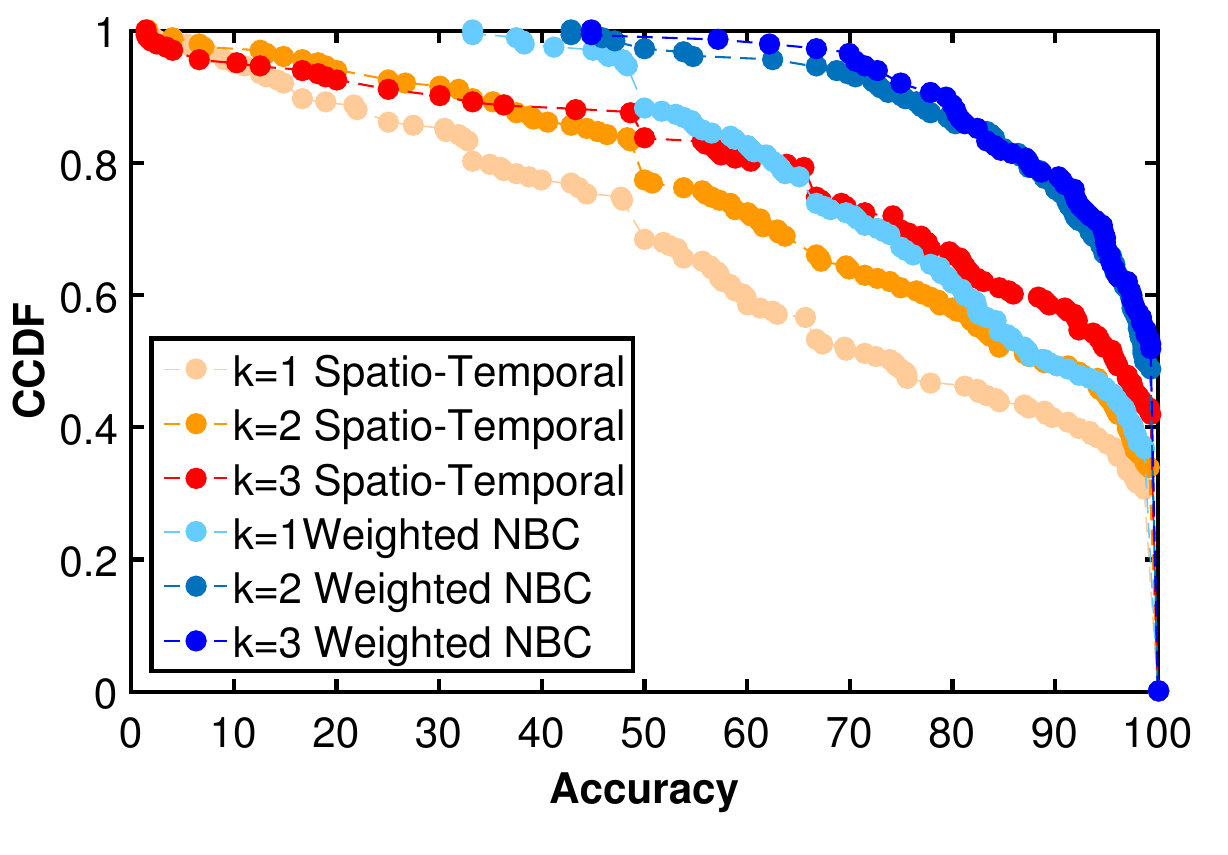}
		\caption{}
		\label{fig:CompareLocationPredictionAccuracy-SP-WeightedBayesian}
	\end{subfigure}
	\begin{subfigure}{0.51\textwidth}
		\includegraphics[width=\textwidth]{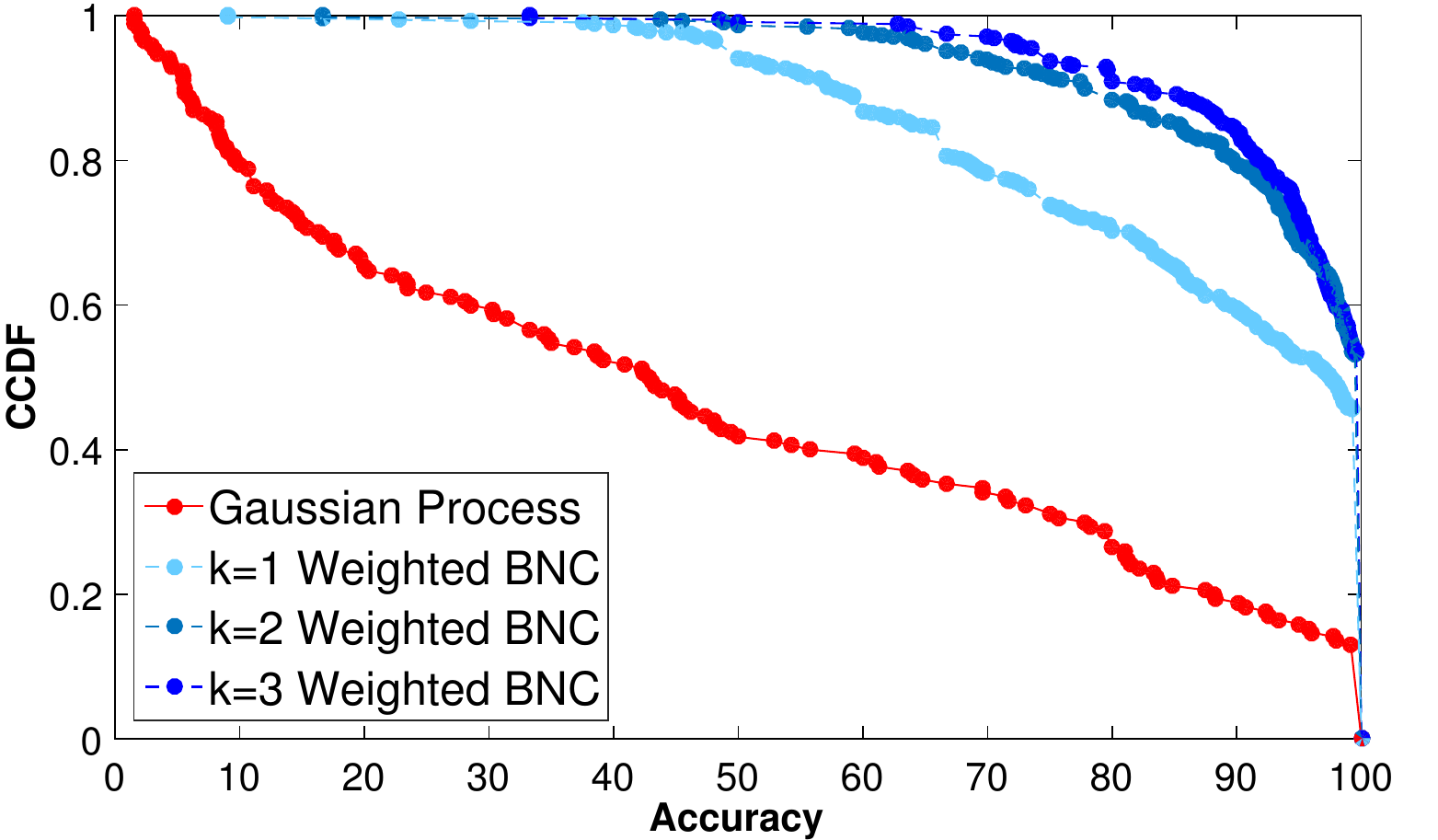}
		\subcaption{}
		\label{fig:CompareLocationPredictionAccuracy-Gaussian-WeightedBayesian}
	\end{subfigure}
	\begin{subfigure}{0.48\textwidth}
		\includegraphics[width=\textwidth]{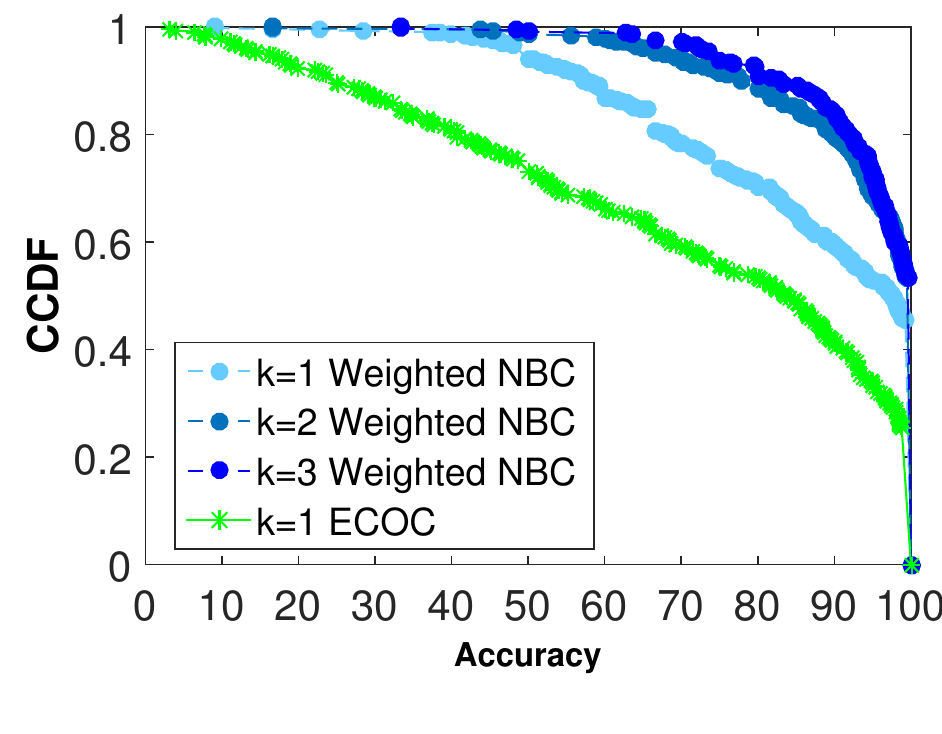}
		\subcaption{}
		\label{fig:CompareLocationPredictionAccuracy-ECOC-WeightedBayesian}
	\end{subfigure}
	\begin{subfigure}{0.51\textwidth}
		\includegraphics[width=\textwidth]{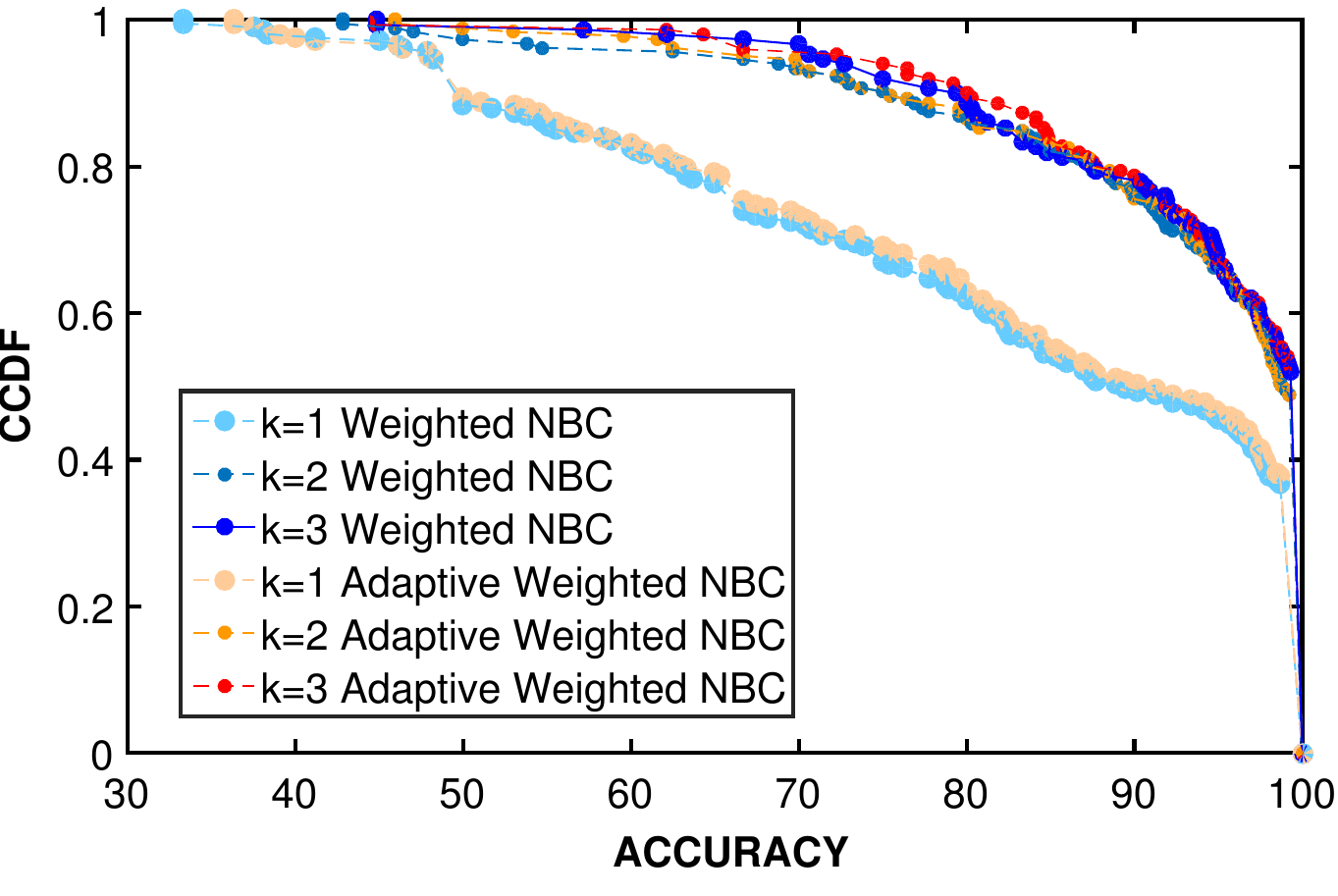}
		\subcaption{}
		\label{fig:Compare AdaptiveCompareLocationPredictionAccuracy}
	\end{subfigure}
	
	\caption{In a) the comparison between our approach and the Gao's solution in terms of PoIs accuracy distribution. In b) the accuracy distribution obtained by the Ciobanu's methods and by our approach. In c) the comparison between the weighted NBC and ECOC version. In d) the comparison between the adaptive weighted NBC and its non-adaptive version.}
	\label{fig:EncounterPredictionComparison}

\end{figure*}

We also compared our approach with two other algorithms: Gao et al. \cite{Gao2012} and Ciobanu et al. \cite{Zhang2015}. Gao's method is based on the i.i.d. (independent and identical distribution)  assumption about spatio-temporal contexts and assume a Gaussian distribution of the daily and hourly temporal contexts. In Figure \ref{fig:CompareLocationPredictionAccuracy-SP-WeightedBayesian} we report the distributions of the accuracy obtained by our approach and Gao's method. By comparing the distributions for $k=1,2,3$ we confirm that our approach outperforms the \cite{Gao2012} in terms of accuracy. For instance, in the $k=1$ case, half of the classifiers trained by the spatio-temporal method obtain an accuracy less than 70\%, while, in our approach, the same level of accuracy is reached by about 80\% of the predictors.The same observation holds for $k=2,3$.

We also compared our approach with the Ciobanu's method. The solution they propose relies on Gaussian processes and focuses on the prediction of the number of encounters, not the encounter PoI. However, Ciobanu's solution is based on the estimation of the same joint distribution adopted by our approach. In the comparison we use Gaussian process (with constant mean function and covariance ARD) to estimate the same underlying joint distribution, changing the conditional event. In Figure \ref{fig:CompareLocationPredictionAccuracy-Gaussian-WeightedBayesian} we report the accuracy distributions obtained by the two methods. We observe that, even for $k=1$, our approach outperforms the Gaussian approach. In fact, the median of the distribution obtained by the Gaussian process predictor is 41\%, while in our approach it reaches about $97\%$.

In another experiment, we compared our Bayesian weighted features predictor approach with ECOC (Error-Correcting Output Code ) predictor which is the multi-class version of SVM classifier. Again here, in Figure \ref{fig:CompareLocationPredictionAccuracy-ECOC-WeightedBayesian}, we can observe that even for $k=1$, our approach outperforms the ECOC approach.

Finally, as mentioned in section \ref{sec_ProbabilisticModel}, feature weights are shared and averaged over all users in the training set. We could increase the level of personalization of the predictor by computing for each user its feature weights. In this case, we obtain an adaptive weighted Bayesian predictor. In the following, we verify whether the increase of the personalization level may result into a better accuracy w.r.t. a non-adaptive predictor. To this aim, Figure \ref{fig:Compare AdaptiveCompareLocationPredictionAccuracy} depicts the comparison among distributions of  the accuracy for weighted and adaptive weighted Bayesian predictors. We observe some minor improvements in the prediction accuracy for the case of adaptive feature weights.

\subsubsection{Encounter duration accuracy}

\begin{figure}
	\centering
	\includegraphics[width=1\textwidth]{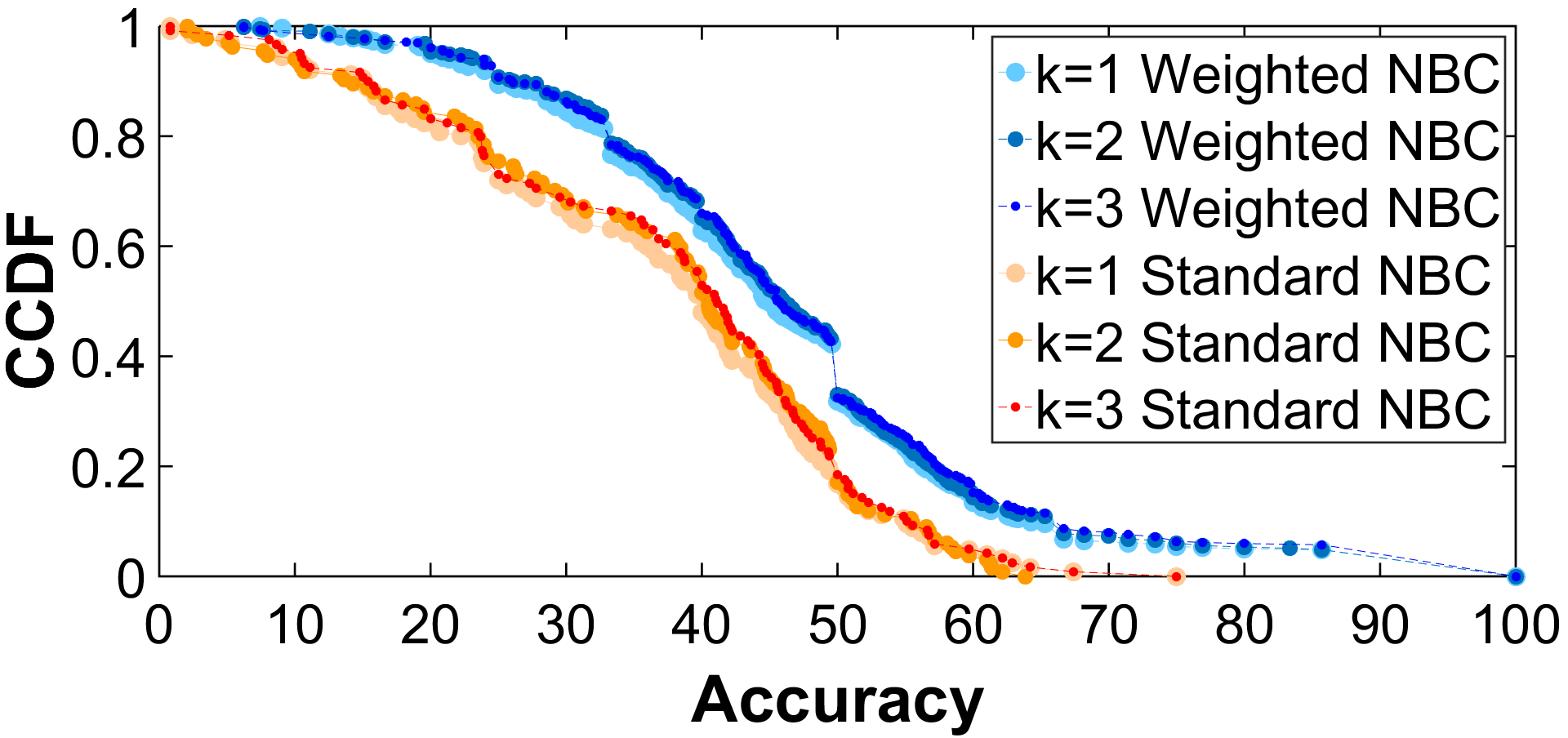}
	\caption{Comparison between the standard NBC and its weighted version in terms of accuracy distribution for the task of the duration prediction.}
	\label{fig:CompareWeighted-UnweightedDurationPredictionAccuracy}
\end{figure}

As shown in Figure \ref{fig:CompareWeighted-UnweightedDurationPredictionAccuracy}, the distribution of the encounter duration prediction accuracy is degraded. In this case, we observe, for each $k$, just around 30\% of classifiers have an accuracy higher than 50\%.

In Figure \ref{fig:CompareWeighted-UnweightedDurationPredictionAccuracy} we compare the distribution of the accuracy for the encounter duration task with standard na\"{i}ve Bayesian and weighted features version predictors. We observe that the feature weighting considerably enhances the accuracy for each $k$.

\subsubsection{Encounter Contacts Predictor}

\begin{figure}
	\centering
	\includegraphics[width=1\textwidth]{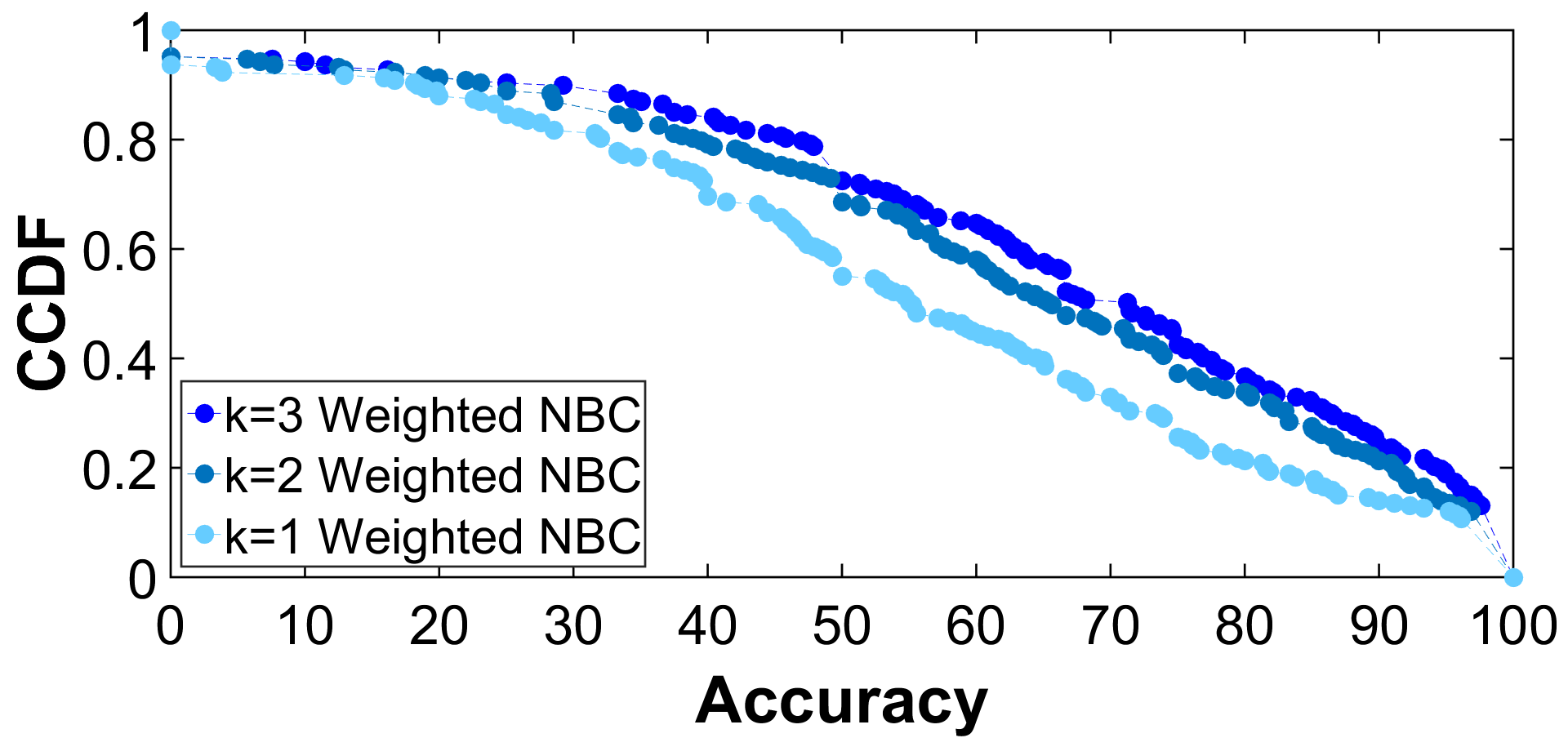}
	\caption{The encounter contacts accuracy distributions for weighted features Bayesian classifier and for $k=1,2,3$.}
	\label{fig:WeightedContactsPredictionAccuracy}
\end{figure}

Finally, we cope with the encounter contacts prediction. The accuracy distributions for the encounter contacts prediction task for weighted features Bayesian classifier are depicted in Figure \ref{fig:WeightedContactsPredictionAccuracy}, for $k=1,2,3$. In this scenario, we observe pronounce differences of accuracy whereas $k$ changes.
For $k=1$ about 20\% of mobile users have accuracy greater than 80\%, while for $k=3$ the same performance level is reached by about 30\% of the population.

\begin{figure}[b!]
	\centering
	\includegraphics[width=1\textwidth]{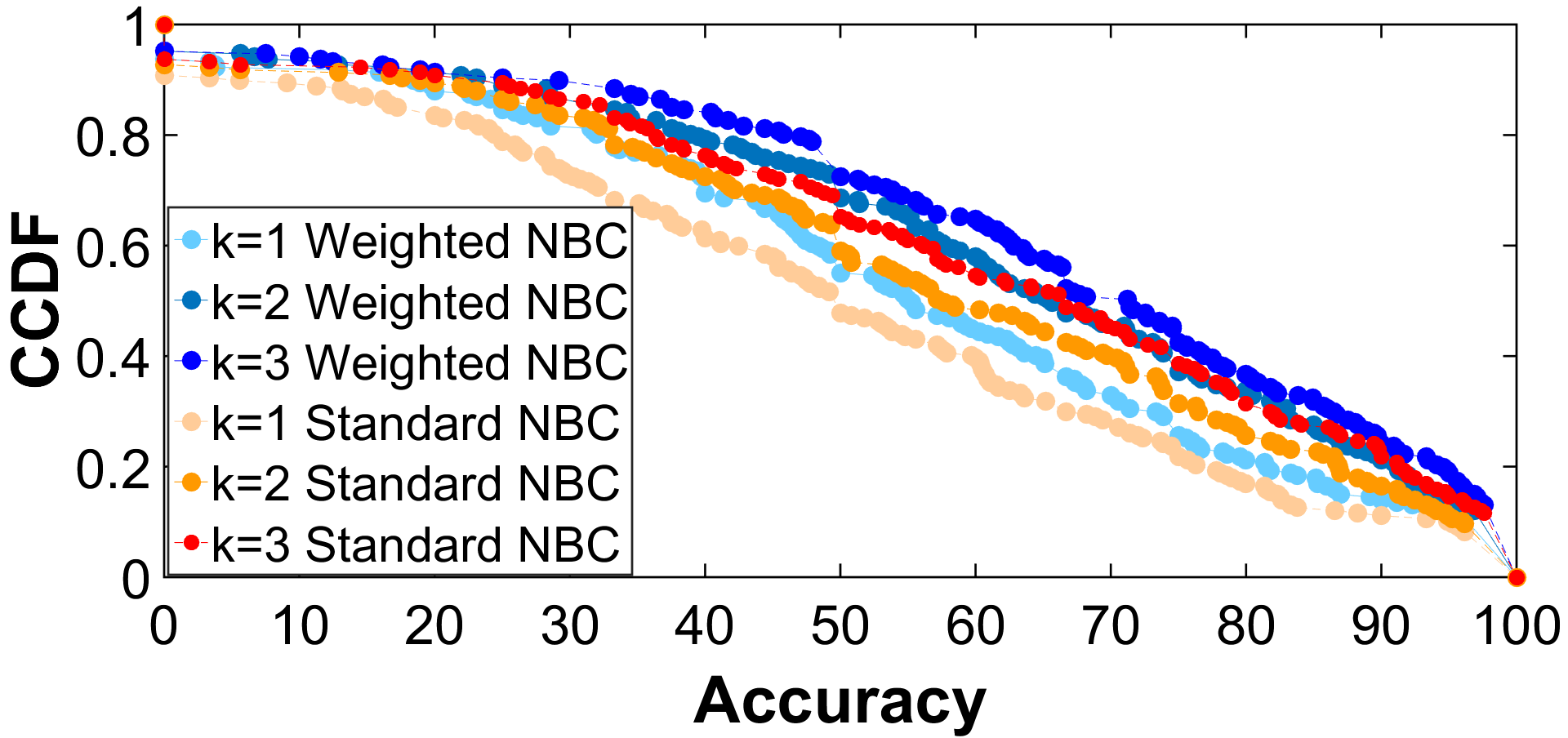}
	\caption{Comparison between the standard NBC and its weighted version in terms of accuracy distributions for the task of the encounter contacts prediction.}
	\label{fig:ComUnWeighted-WeightedContactsPredictionAccuracy}
\end{figure}

In Figure \ref{fig:ComUnWeighted-WeightedContactsPredictionAccuracy} the accuracy distributions for the encounter contacts prediction achieved by standard  na\"{i}ve Bayesian predictor and its weighted features version, are depicted. The enhancements of the accuracy are significant in the weighted cases for $k=1,2$. 

\subsection{Colocation prediction}

In colocation trace we focus on a subset of users with at least 350 records, to have enough data for training classifier. In this section, we evaluate the colocation prediction accuracy performance for PoIs, and also contacts for input temporal query.

\subsubsection{Colocation PoI prediction}

\begin{figure}
	\centering
	\includegraphics[width=1\textwidth]{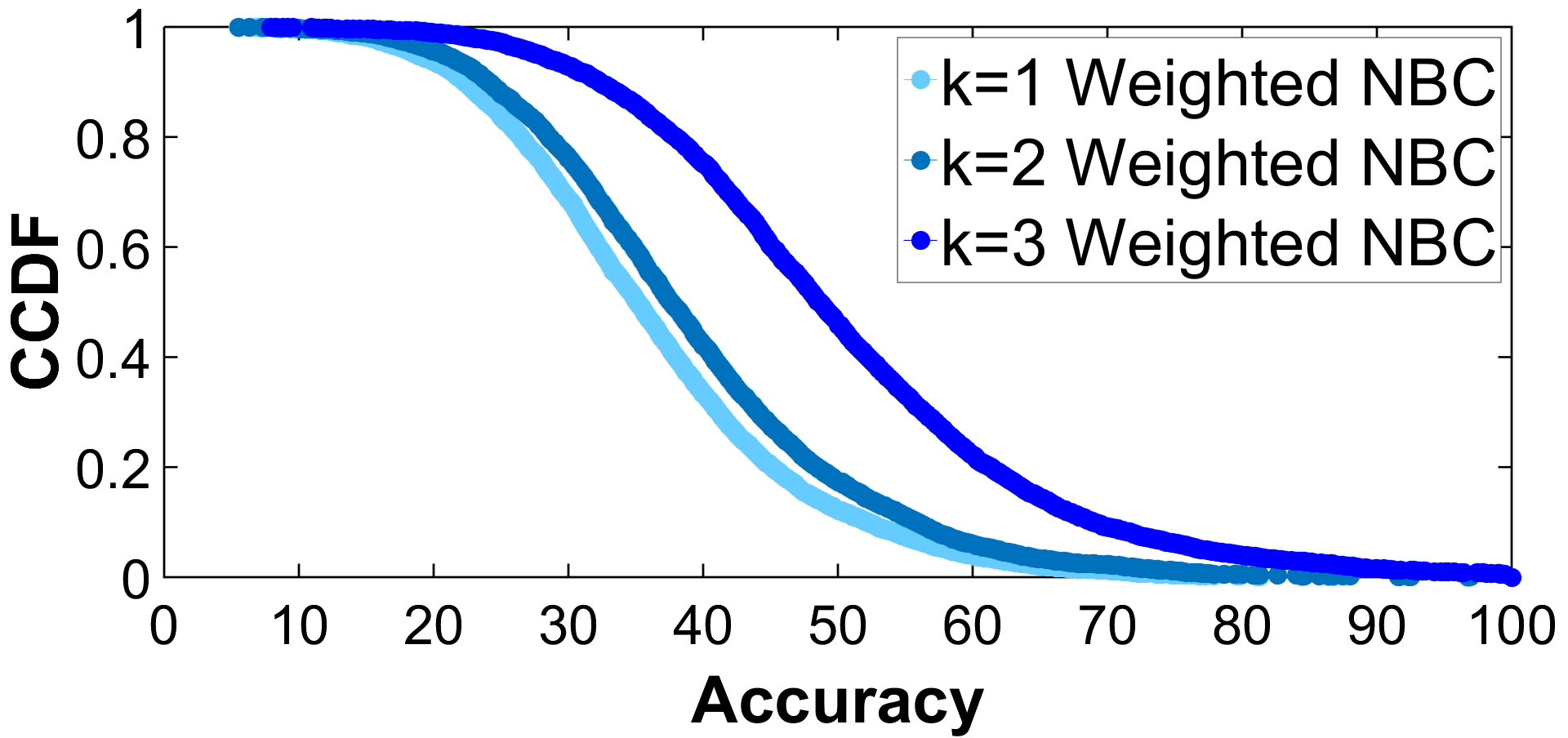}
	\caption{The distribution of the PoIs colocation accuracy for the Bayesian classifier with weighted features.}
	\label{fig:ColocationLocationDistribution-p-123-Weighted}
\end{figure}

In the colocation trace, in Figure \ref{fig:ColocationLocationDistribution-p-123-Weighted} when $k=3$, around 50\% of mobile users, have more than 50\%  accuracies for PoI prediction. For $k=1$ and $k=2$, we observe that around just 15\% and 20\% of predictors obtain more than 50\% accuracies for the task.

\begin{figure}
	\centering
	\includegraphics[width=1\textwidth]{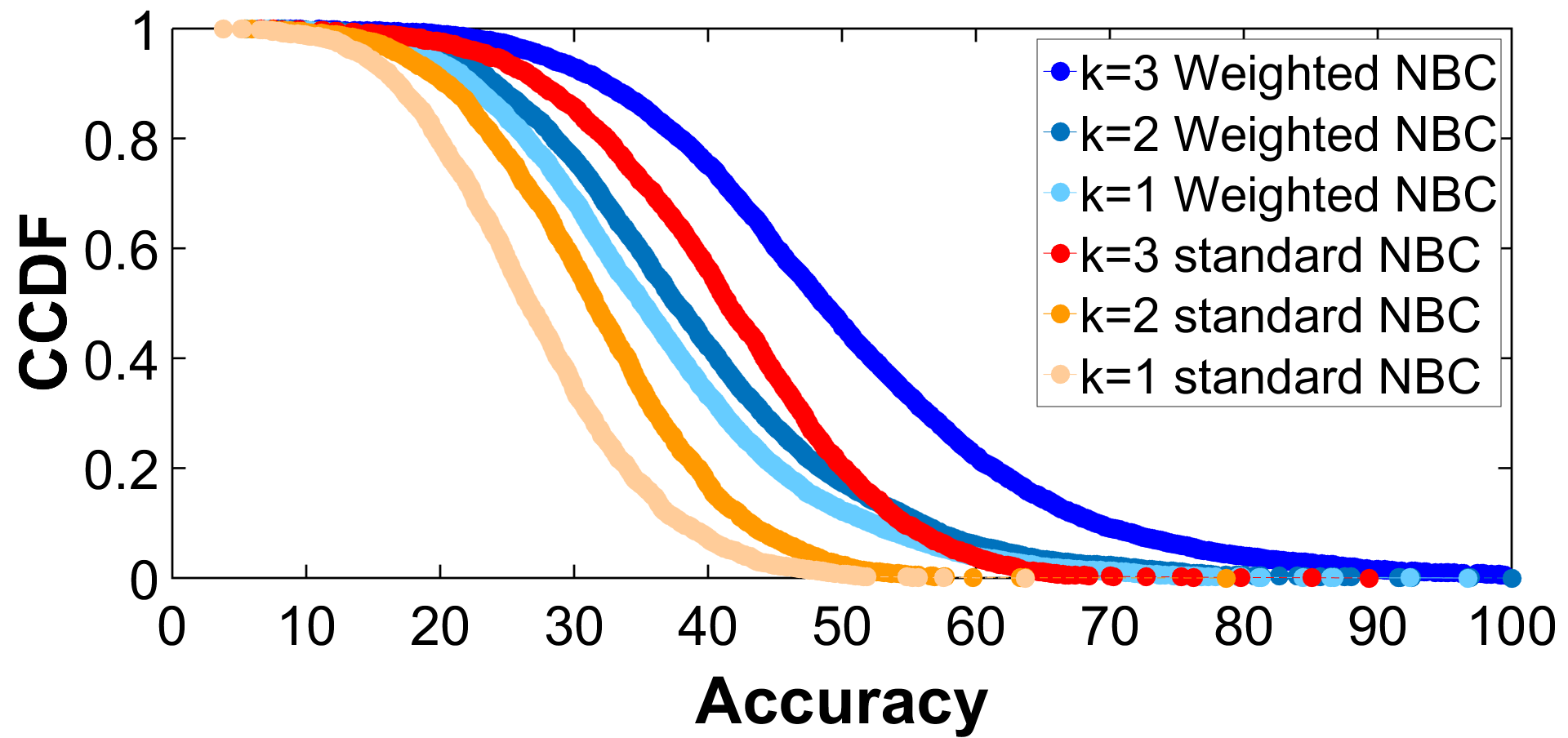}
	\caption{Comparison between the standard na\"{i}ve Bayesian classifier and its weighted version in terms of PoIs colocation accuracy distribution.}
	\label{fig:CompareCoLocationPredictionAccuracy}
\end{figure}

In Figure \ref{fig:CompareCoLocationPredictionAccuracy} the accuracy distributions for the colocation PoIs prediction achieved by Bayesian with weighted features and standard  na\"{i}ve Bayesian (NBC) predictors are depicted. The enhancements of the accuracy are significant in the weighted cases for all values of $k$. 

By comparing the accuracy in both encounter and colocation, we observe that the PoI colocation predictors are less accurate than those trained on the encounter trace, most probably due to the higher spatio-temporal sparsity of CDR dataset w.r.t. WiFi dataset.
For colocation scenario, when $k=3$, around 50\% of mobile users, have more than 50\% accuracies. 

\subsubsection{Colocation contact accuracy}

\begin{figure}
	\centering
	\includegraphics[width=1\textwidth]{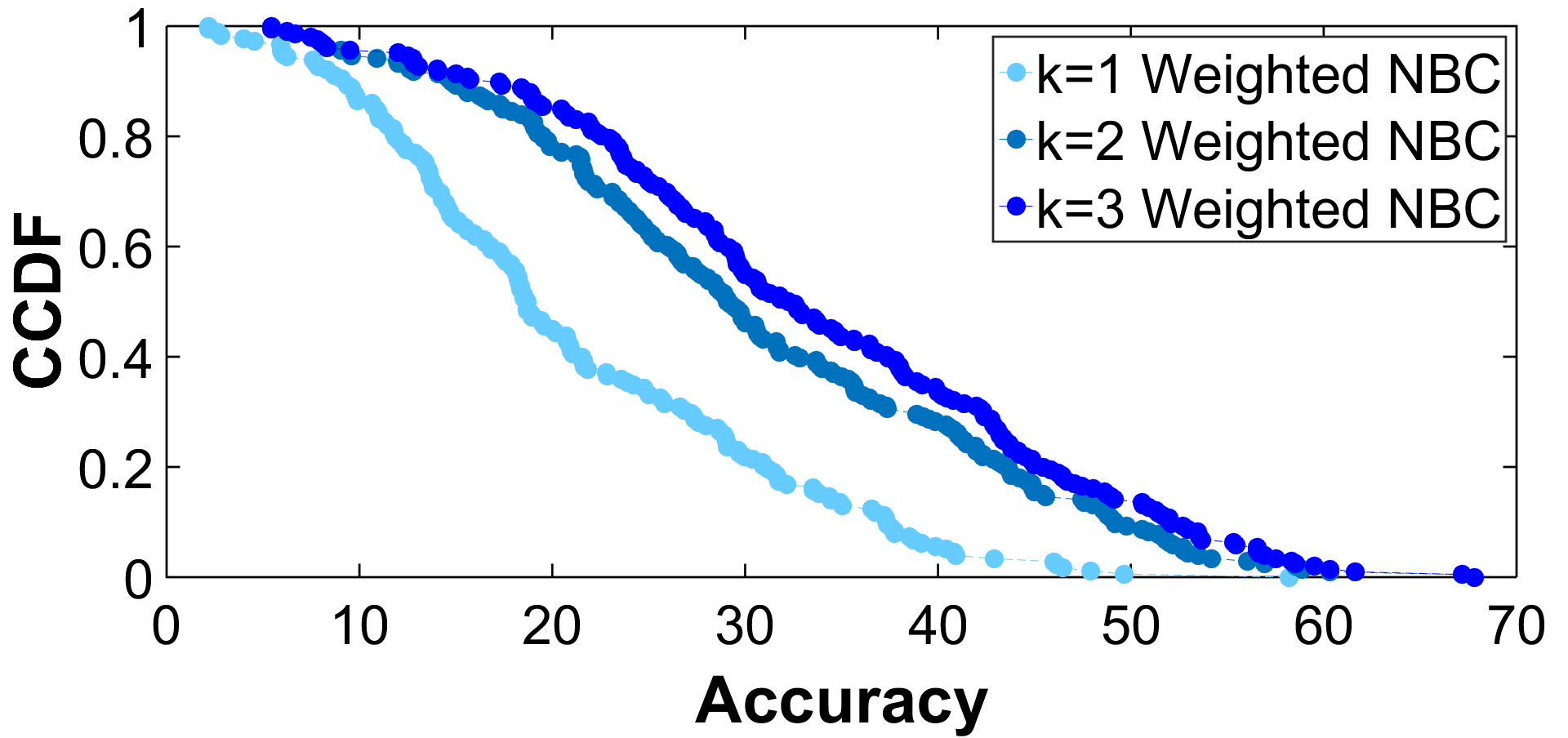}
	\caption{The accuracy distribution for the colocation contact prediction for $k=1,2,3$.}
	\label{fig:WeightedColocContactsPredictionAccuracy}
\end{figure}

In the colocation scenario, the accuracy distributions for Bayesian with weighted features, shown in Figure \ref{fig:WeightedColocContactsPredictionAccuracy}, points out worse performances w.r.t. the encounter scenario. None of the predictors have obtained an accuracy greater than 70\%.
We suppose that the performance gap between the two scenarios may be due to the larger coverage area of the cellular tower, the temporal sporadicity w.r.t. the WiFi APs and to a higher spatio-temporal regularity of the students in a campus with respect to the people in a metropolitan area.

\section{Conclusion}\label{sec:conclosion}

In this paper, we cope with the prediction of the encounter and colocation events and some of their properties, such as  PoIs, duration and also the people involved in the meeting. Being able to forecast these properties by simply leveraging the mobility patterns of the user, is one of the crucial aspects in applications directly relying on the human behaviors, e.g. opportunistic networks, online and location-based social networks or epidemiology.

To reach our goal, we turn the main task into a multi-class classification problem. Specifically, we use spatio-temporal mobility information to train multi-class weighted features Bayesian classifier which predicts with high accuracy the next encounter and colocation events along with their characteristics. To mitigate the effects of the conditional independence assumptions in the na\"{i}ve Bayesian, we introduced the feature weighting based on the Kullback-Leibler divergence.

We evaluate our approach on two different datasets which have been extracted from large-scale WiFi and CDR datasets. As for the encounter prediction task, the classifier has obtained a good accuracy for most of the users, while in the colocation trace, the performances worsen due to the high spatio-temporal sparsity and coarseness of the users' movements in CDR dataset. 

As regards, the prediction of the people involved in an encounter event, half of the predictors have obtained an accuracy greater than 70\%, whereas for colocation events the performance considerably worse.
In general, we observed the higher accuracy for encounter events prediction w.r.t. colocation events. Since encounter events are inherently regular in campus environments, participants (students, staffs, and professors) have the regular schedule during working days, means usually they meet in the same class almost at the same hours of the day and days of the week, during all academic weeks. So spatio-temporal regularity of events in encounter trace will be higher w.r.t. colocation traces extracted from CDR datasets collected from metropolitan area.

Finally, we also compare our methods against some states of -the art approaches presented in the literature. In all cases, the weighted feature Bayesian predictor outperforms the other prediction algorithms.

 \section{References}\label{sec:references}
 \bibliographystyle{abbrv}

 \bibliography{matteoBib}

\end{document}